\documentclass[conference]{IEEEtran}
\IEEEoverridecommandlockouts
\usepackage{cite}
\usepackage{amsmath,amssymb,amsfonts}
\usepackage{algorithmic}
\usepackage{graphicx}
\usepackage{textcomp}
\usepackage{xcolor}

\usepackage{array}
\usepackage{booktabs}
\usepackage{tabularx}
\usepackage{multirow}

\usepackage{algorithm}
\usepackage{mathtools}
\usepackage{amsmath}

\begin{document}

\title{SWDPM: A Social Welfare-Optimized Data Pricing Mechanism}


\author{
\IEEEauthorblockN{Yi Yu}
\IEEEauthorblockA{\textit{Urban Computing Lab} \\
\textit{Shanghai AI Laboratory}\\
Shanghai, China \\
yuyi@pjlab.org.cn}
\and
\IEEEauthorblockN{2\textsuperscript{nd} Shengyue Yao}
\IEEEauthorblockA{\textit{Urban Computing Lab} \\
\textit{Shanghai AI Laboratory}\\
Shanghai, China \\
yaoshengyue@pjlab.org.cn}
\and
\IEEEauthorblockN{3\textsuperscript{rd} Juanjuan Li}
\IEEEauthorblockA{\textit{The State Key Laboratory of Management and Control for}\\
\textit{Complex Systems, Institute of Automation,} \\
\textit{Chinese Academy of Sciences}\\
Beijing, China \\
juanjuan.li@ia.ac.cn}
\and
\IEEEauthorblockN{4\textsuperscript{th} Fei-Yue Wang}
\IEEEauthorblockA{\textit{The State Key Laboratory of Management and Control for}\\
\textit{Complex Systems, Institute of Automation,} \\
\textit{Chinese Academy of Sciences, Beijing, China}\\
\textit{The Macau Institute of Systems Engineering,}\\
\textit{Macau University of Science and Technology, Macau, China} \\
feiyue.wang@ia.ac.cn}
\and
\IEEEauthorblockN{5\textsuperscript{th} Yilun Lin}
\IEEEauthorblockA{\textit{Urban Computing Lab} \\
\textit{Shanghai AI Laboratory}\\
Shanghai, China \\
linyilun@pjlab.org.cn}
}

\maketitle

\begin{abstract}
Data trading has been hindered by privacy concerns associated with user-owned data and the infinite reproducibility of data, making it challenging for data owners to retain exclusive rights over their data once it has been disclosed. Traditional data pricing models relied on uniform pricing or subscription-based models. However, with the development of Privacy-Preserving Computing techniques, the market can now protect the privacy and complete transactions using progressively disclosed information, which creates a technical foundation for generating greater social welfare through data usage. In this study, we propose a novel approach to modeling multi-round data trading with progressively disclosed information using a matchmaking-based Markov Decision Process (MDP) and introduce a Social Welfare-optimized Data Pricing Mechanism (SWDPM) to find optimal pricing strategies. To the best of our knowledge, this is the first study to model multi-round data trading with progressively disclosed information. Numerical experiments demonstrate that the SWDPM can increase social welfare 3 times by up to 54\% in trading feasibility, 43\% in trading efficiency, and 25\% in trading fairness by encouraging better matching of demand and price negotiation among traders.

\end{abstract}


\section{Introduction}
In the new era of digitalization, ubiquitous data with substantial production value are generated, which leads to the rapid development of new industries such as data collaboration platforms and Data as a Service(DaaS)\cite{zhengServiceGeneratedBigData2013}. Especially, the big data analytics market size reaches USD 271.83 billion in 2022 while predicted to reach USD 655.53 billion by 2029\cite{insights_big_2022}, and its development is expected to improve social welfare by bringing value increments to data suppliers, buyers, and the entire society\cite{tripathiDoesInformationCommunications2020}.

Governments, markets, and academic communities have a growing awareness of the data value, meanwhile conducting numerous data pricing practices and studies from different perspectives. In most government-led platforms, data are freely shared for public use\cite{labCityDataExchange2017}. In data markets, data were mainly traded with uniform or subscription-based pricing for the purpose of maximizing revenue\cite{agarwalMarketplaceDataAlgorithmic2019, zhangSurveyDataPricing2020, chenSellingDataMachine2022}. In academic fields, various data pricing models were studied from value-based, negotiation-based, and cost-based aspects, applying methodologies in economics, marketing, machine learning, etc\cite{yuDataPricingStrategy2017, tianDataBoundaryData2022, koutrisQueryBasedDataPricing2015}. However, it is worth noticing that current data pricing models are mainly proposed on the assumption that all information would be disclosed once traded, thus data is likely to be only traded once or multi-times homogeneously\cite{peiSurveyDataPricing2020,yuDataPricingStrategy2017,koutrisQueryBasedDataPricing2015, tianDataBoundaryData2022}. In this paper, we call it a complete disclosure assumption.

The development of privacy-preserving technologies supports  multi-round data trading by quantifying information disclosure and improving reliability during the trading\cite{niuMakingBigMoney2019,zhouBriefSurveyAnonymization2008,liTheoryPricingPrivate2015, mothukuriSurveySecurityPrivacy2021, liNovelFrameworkData2022}. 
Since data contains a vast amount of information, it is possible to trade the data over multiple rounds by ensuring that only part of the information is disclosed in each trading, thus transforming data trading from a single, non-sequential process to a series of temporal-related trading. By adopting the multi-round trading approach, traders can leverage the correlation between trading to achieve their target information and revenue, which expands the application range and unlocks the potential value of data by maximizing its utility. However, existing models like uniform or subscription-based pricing can hardly meet these needs, as they mostly lack the capture of dynamics or consideration of sustainability due to the complete disclosure assumption\cite{peiSurveyDataPricing2020}. 

\begin{figure}
    \centering
    \includegraphics[width=1\linewidth]{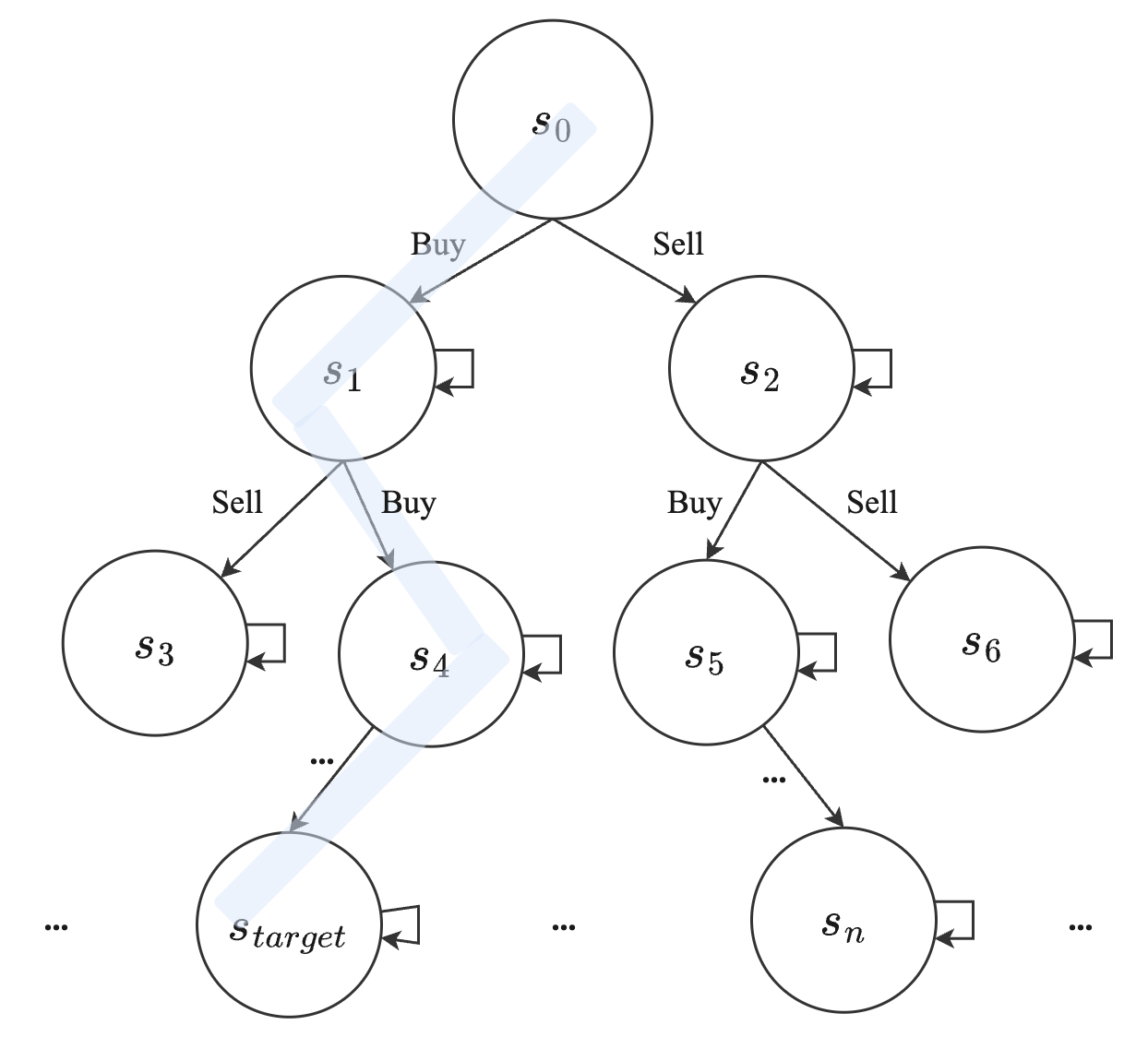}
    \caption{Data trading as Markov decision process}
    \label{decision_process}
\end{figure}

In the multi-round trading pattern, the data trading can be modeled as a multi-agent Markov Decision Process (MDP), where each trader's state can be characterized by information volume and currency, and action can be characterized by buying or selling information or maintaining, as shown in Fig. \ref{decision_process}. The trader's current state is only related to the previous state and action, showing Markov properties\cite{changIncorporatingMarkovDecision2017}, and the state transition depends on its own actions and others' actions, making it a mixed-strategy game. In this process, pricing plays a crucial role in deciding the exchange ratio of information disclosure and currency, which is related to value and volatility. Reasonable pricing can incentivize traders to participate and unlock the data value, generating surplus benefits for all buyers and sellers, which ultimately increases social welfare that equals the sum of surpluses\cite{jiangHighspeedRailPricing2021}.

Upon modeling the data trading process as MDP, we propose a social welfare-optimized data pricing mechanism (SWDPM) that consists of a matchmaking method and employs reinforcement learning (RL) to effectively address the multi-round trading pattern. Instead of directly calculating value and volatility, the RL-based mechanism could optimize pricing strategies by learning from the interactions between traders, thereby incentivizing the trading. Numerical simulations are employed to validate the feasibility of the proposed mechanism. Results demonstrate more positive collaboration among traders, and superiority in feasibility, efficiency, fairness, and social welfare increment compared with traditional pricing mechanisms. 

With the aforementioned research gaps, the key contributions of this paper are: (1) the data trading process is modeled by a matchmaking-based MDP, which enables capturing the pricing dynamics in the multi-round trading pattern, as well as traders' characteristics; (2) an RL-based social welfare-optimized trading mechanism is proposed to enhance the data trading's sustainability and fully unlock the data value by maximizing the social welfare.   

\section{Related Work}
In the data pricing field, researchers conduct studies mainly from two aspects: designing exclusive and sophisticated models that can estimate the volatility through trader's negotiation; applying economic pricing methods that can estimate the data value from value-based and cost-based perspectives. Regarding game theory-based models, there are uniform, subscription-based, and negotiation-based data pricing\cite{wuOptimalFixedUpToPricing2019, linFullyDecentralizedInfrastructure2020, agarwalMarketplaceDataAlgorithmic2019, zhangSurveyDataPricing2020, chenSellingDataMachine2022, liNovelGSPAuction2022, liNovelGSPAuction2019}. Because of the complete disclosure assumption which implies that data is likely traded only once, existing negotiation-based pricing models have defects in capturing the dynamic characteristics or generalizing their optimal solutions to more scenarios\cite{peiSurveyDataPricing2020}. Most models also have defects in feasibility because they are too complex to be applied in reality.
As for economic theory-based models, there are various value-based and cost-based data pricing models, such as data quality-based\cite{yuDataPricingStrategy2017}, Shapley value-based\cite{tianDataBoundaryData2022}, query-based\cite{koutrisQueryBasedDataPricing2015}, and quantity-based models\cite{mehtaHowSellDataset2019}. However, a key challenge persists in building a sustainable and incentive pricing mechanism, which is desired in the era of digitalization\cite{fernandezDataMarketPlatforms2020}. 

In the broader pricing research, MDP and RL have received widespread attention for the mechanism design recently. Multiple fields, such as energy markets, insurance industries, and sponsored search auctions, regard pricing problems as a series of decision-making problems\cite{maestreReinforcementLearningFair2019, kongOnlinePricingDemand2020, xuDeepReinforcementLearning2019, krasheninnikovaReinforcementLearningPricing2019, shenReinforcementMechanismDesign2020}. The trading process can be modeled as MDPs or constrained MDPs\cite{xuDeepReinforcementLearning2019, krasheninnikovaReinforcementLearningPricing2019}, and various algorithms like Q-learning\cite{maestreReinforcementLearningFair2019}, deep deterministic policy gradient (DDPG)\cite{xuDeepReinforcementLearning2019}, long short-term memory(LSTM)\cite{kongOnlinePricingDemand2020} are introduced to support the construction of pricing mechanism. Many numerical simulations and experimental validation demonstrate that RL-based mechanisms show high performance and robustness in solving complex pricing problems, achieving  fairness and revenue maximum by setting proper objectives\cite{maestreReinforcementLearningFair2019, krasheninnikovaReinforcementLearningPricing2019}. 

Major data pricing research lacks the consideration of dynamics and sustainability because of the complete disclosure assumption, while in broader pricing research, the dynamics and sustainability of pricing mechanisms are achieved with the application of MDP and RL. With the support of privacy-preserving technologies, it is possible to design a pricing mechanism for data trading scenarios using MDP and RL that can accurately describe the trading process and provide incentive pricing results without losing generality, thereby unlocking the data value. Various objectives such as fairness, sustainability, and social welfare could be taken into account in these mechanisms. 

\section{Problem Statement}

In this research, we focus on unlocking the data value to enhance social welfare through multi-round trading between data traders. Given this purpose, the data trading process is modeled as a matchmaking-based MDP, which is elaborated in this section. The variables and parameters used to propose the SWDPM are summarized in Table. \ref{notation_table}.
\begin{table}[htbp]
\centering
\caption{Notations of variables and parameters}
\begin{tabular}{ll}
\toprule
Notation & Meaning\\
\midrule
$\mathcal{T}$ & Collection of traders \\
$s_t^i$  & State of trader $i$ at time $t$ \\
$S_t$ & Collection of states at time $t$ \\
$vt^i$ & Target information volume of trader $i$\\
$v_t^i$  & Information volume of trader $i$ at time $t$\\
$c_t^i$  & Currency amount of trader $i$ at time $t$  \\
$a_t^i$ & Action taken by trader $i$ at time $t$\\
$A_t$ & Collection of actions at time $t$\\
$dv_t^i$ & Volume of disclosed information by trader $i$ at time $t$ \\
$dc_t^i$  & Amount of Changed currency by trader $i$ at time $t$ \\
$I_i$ & Trading acceptance intention of trader $i$ \\
$I$  & Collection of trading acceptance intentions\\
$dc_{min}$  & Minimum price \\
$\Delta A_t$ & Unsolved action set at time $t$ \\
$\Delta a_t$ & Unsolved action in $\Delta A_t$ \\
$\Delta dv_t^i$ & Volume of untraded disclosed information \\
$p_a$ & Probability of state transfer by taking $a$ \\
$r_t^i$  & Reward of trader $i$ at time $t$ \\
$R_t$  & System reward at time $t$ \\
$w(a_t^i)$  & Trader's surplus by taking $a_t^i$ \\
$W$ & Social welfare increment \\
$Q$ & Action-value function \\
$Q^*$  & Optimal action-value function \\
$\Tilde{Q}$ & Q table \\
$A_t^*$ & Optimal collection of actions at time $t$ \\
$\mathcal{D}$  & Historical trading data \\
$\eta$ & Standard price for one unit disclosed information \\
$\delta$ & Discount rate \\
$\theta$  & Weighting factor of fairness \\
$\lambda$  &  Weighting factor of the penalty \\
$\alpha$ &  Learning rate  \\
$\gamma$ & Discount factor  \\
$\xi$ &  Max episodes \\
$\Phi_{f}$ & Evaluation metric of feasibility\\
$\Phi_{e}$  & Evaluation metric of effectiveness\\
$\Phi_{r}$  & Evaluation metric of fairness\\
\bottomrule
\end{tabular}
\label{notation_table}
\end{table}

\subsection{Data trading as a matchmaking based Markov Decision Process} 
The MDP is widely used in representing stochastic decision-making processes in dynamic systems; specifically, the system dynamics can be presented as the evolution of set $(s,a,p_{a},r_{a})$, where $s$ is the system state, $a$ is the action taken at $s$, $p_{a}=Pr\{s_{t+1}=s'|s_{t}=s,a_{t}=a\}$ is the probability that the system state transfers to $s'$ at time $t+1$ while at $s$ and taking $a$ at time $t$, whereas $r_a$ is the reward function of taking $a$ at $s$. The set $(s,a,p_{a},r_{a})$ in the data trading process is defined based on assumptions as follows:

\begin{itemize}
    \item[(1)] The volume of disclosed information from the raw data can be measured and partitioned into multiple portions, based on which multi-round data trading is theoretically and practically feasible.
    \item[(2)] Traders can sell or buy data with multiple traders per time step, while one agent can only sell or buy data from another agent one time per time step.
    \item[(3)] The trading process for a trader is regarded as a uni-directional process. Specifically, traders only consider buying (selling) data while the information volume that they possess is less (greater) than their target, respectively. 
    \item[(4)] Both the information volume and the currency that is used to trade data are discretized with the minimum units $u_v$ and $u_c$, respectively. 
\end{itemize}

According to these assumptions, for a group of traders $\mathcal{T}=\{1,2,...,N\}$ with $N$ traders, the definition of MDP, including the definition of set $(s,a,p_{a},r_{a})$ and the matchmaking method, is elaborated as follows:

\paragraph{State}
For trader $i$ in $\mathcal{T}$, its state at time step $t$, $s^{i}_{t}$ can be represented as a vector of $[vt^{i},v_{t}^{i},c_{t}^{i}]$; where $v_{t}^{i} \in \mathbb{R}_{\geq 0}$ represent the accumulated volume of information that is disclosed from the raw data, which is possessed by trader $i$ at time step $t$, and $vt^{i} \in \mathbb{R}_{\geq 0}$ represents the target accumulated information volume that trader $i$ expects to possess. In addition, $c_{t}^{i} \in \mathbb{R}_{\geq 0}$ represents the amount of currency that is possessed by trader $i$ at time step $t$. Note that $vt^{i}$ is a fixed value that is decided before the trade process begins. In addition, according to assumption (3), while $v_{t}^{i} < vt^{i}$, trader $i$ seeks to buy data by spending its currency $c_{t}^{i}$ until $v_{t}^{i} = vt^{i}$; similarly, trader $i$ will keep selling data while $v_{t}^{i} > vt^{i}$ until its $v_{t}^{i}$ reaches the $vt^{i}$. 

\paragraph{Action}
Trader $i$ transfers between states by taking action $a_t^i$, which can be represented as a vector of $[dv_{t}^i, dc_{t}^i]$; where $dv_t^i \in [-v_{t}^{i},\infty)$ and $dc_t^i \in [-c_{t}^{i},\infty)$ are the accumulated disclosed information volume and accumulated changed currency, respectively, during trading that trader $i$ takes at time step $t$. In addition, the action $a_t^i$ can be regarded as a price proposal of selling (buying) a certain amount of disclosed information for trader $i$, thus a feasible proposal should comply with the trading acceptance intention of trader $i$. In this research, the trading acceptance intention is represented by $I_i$, which is formulated by (\ref{intention}):

\begin{subequations}
\label{intention}
\begin{equation}
I_i(dv_{t}^i,dc_{t}^i) = \begin{cases}
  1,              & dc_{t}^i(j) \geq dc_{min}(dv_t^i) \\
  0,                    & otherwise
\end{cases}  
\end{equation}
where
\begin{equation}
dc_{min}(dv_t^i) \underrightarrow{R} \begin{cases}
  [-\eta \cdot dv_t^i, - (1-\delta) \cdot \eta  \cdot dv_t^i],              & dv_{t}^i(j) \geq 0 \\
  [-\eta \cdot dv_t^i, -(1+\delta) \cdot \eta \cdot dv_t^i],                    & dv_{t}^i(j) <  0
\end{cases}    
\end{equation}
\end{subequations}
where $\eta$ is a constant reflecting the standard price for one unit portion of disclosed information $u_v$ and $\delta \in (0,1)$ is the discount rate. In general, if the price ($dc_t^i$) proposed for the disclosed information ($dv_t^i$) is greater than the minimum price $dc_{min}$, the proposal can be accepted by trader $i$, that $I(i)=1$. Specifically, while trader $i$ is buying data, a lower price is expected with a greater portion of the disclosed information volume, thus $dc_{min}$ is randomly selected from the set $[-\eta \cdot dv_t^i, - (1-\delta) \cdot \eta \cdot dv_t^i]$; on the contrary, a higher price is expected while trader $i$ is selling a greater portion of the disclosed information volume, thus $dc_{min}$ is randomly selected from set $[-\eta \cdot dv_t^i, -(1+\delta) \cdot \eta \cdot dv_t^i]$.


\paragraph{Matchmaking method}
Given the collection of actions for traders in $\mathcal{T}$, $A_t=\{a_t^1,...a_t^N\}$, and their trading acceptance intentions $I=\{I_1,\ldots,I_N\}$, trading between traders are executed automatically according to the matchmaking method, which is a wide-spread maker/taker model in the equity market \cite{harrisMakertakerPricingEffects2013}. The matchmaking method is described by algorithm \ref{matchmaking}.

\begin{algorithm}[h]
\caption{Matchmaking method by the maker/taker model}\label{matchmaking}
\begin{algorithmic}[1]
\REQUIRE Action set $A_t$, trading acceptance intention set $I$
\ENSURE Unsolved action set $\Delta A_t=\{\Delta a_{t}^1,\Delta a_{t}^2,...,\Delta a_{t}^N\}$
\STATE Initialization $\Delta A_t = A_t$, $m=\infty$
\WHILE{$\Delta A \neq \varnothing$ and $m \neq 0$}
    \STATE Sort $\Delta A_t$ with a random order
    \STATE m=0
    \FORALL{$\Delta a_t$ in $\Delta A_t$}
        \STATE get trader index of $\Delta a_t$ as i
        \IF{$dv^i_t = 0$ and $dc^i_t = 0$}
            \STATE $\Delta A$ = $\Delta A / \{\Delta a_t\}$
            \STATE \textbf{continue}
        \ENDIF
        \FORALL{$\Delta a_t'$ in $\Delta A_t/\{\Delta a_t\}$}
            \STATE get trader index of $\Delta a_t'$ as j
            \IF{$dv^j_t \cdot dv^i_t <0$ and $|dv^j_t| \leq |dv^i_t|$ and $|dc^j_t| \leq |dc^i_t|$ and $I_{i}(-dv^j_t,-dc^j_t)=1$}
                \STATE $\Delta a_t=(dv^i_t+dv^j_t,dc^i_t+dc^j_t)$
                \STATE m=m+1
                \STATE \textbf{break}
            \ENDIF
        \ENDFOR
    \ENDFOR
\ENDWHILE  
\end{algorithmic}  
\end{algorithm}

\paragraph{State transfer probability}
According to the matchmaking method in algorithm \ref{matchmaking}, given the collection of states of traders in $\mathcal{T}$, $S_t$, and the collection of actions, $A_t$, trader $i$ can transfer from $s_t$ to $s_{t+1}$ only when all trading successfully executed, that $\Delta A_t = \varnothing$, thus $p_a=\prod _ {i \in N} Pr\{s_{t+1}^i|s_{t}^i,a_t^i\}$; otherwise, the trading proposals fail and all traders stay in $S_t$. Although the trading process is sophisticatedly modeled and intricate to get a perfect knowledge of $p_a$, the action can be optimized through the Monte-Carlo method.       

\paragraph{Reward}
The objective of SWDPM is to optimize social welfare, and the reward function is designed accordingly with trader's surplus \cite{jiangHighspeedRailPricing2021}, trading fairness \cite{agarwalMarketplaceDataAlgorithmic2019}, and success rate:
\begin{subequations}
\label{reward_function}
\begin{equation}
    r_t^i= w(a_t^i)+ \theta f(a_t^i) + \lambda |\Delta dv_t^i|
\end{equation}
where
\begin{equation}
    w(a_t^i)=|dc_t^i|-|dc_{min}(dv_t^i)|
\end{equation}
demonstrates the trader's surplus by taking $a_t^i$, and 
\begin{equation}
    f(a_t^i)=\frac{dc_t^i}{dv_t^i}-\frac{\sum_{i \in N}dc_t^i}{\sum_{i \in N}dv_t^i}
\end{equation}
\end{subequations}
demonstrates fairness by taking $a_t^i$; $\theta$ is the weighting factor of fairness; whereas $\lambda \Delta dv_t^i$ represents the penalty of unsuccessful trading, where $\Delta dv_t^i$ represents the volume of disclosed information that is not traded through the matchmaking method, which can be found in $\Delta A_t$, and $\lambda$ is the weighting factor of the penalty.

\subsection{Data pricing using Q-learning} 
Given the matchmaking-based MDP of the data trading process, the optimal pricing can be solved via deriving the optimal action-value function $Q^*$. In this research, a centralized learning and centralized execution scheme is adopted, thus the optimal system action-value function $Q^*(S_t, A_t)$ is derived through a classical Q-learning method.

In general, the optimal collection of actions is obtained through (\ref{optimal_pi}):
\begin{equation}
    A^*_t=\operatorname*{argmax}\limits_{A} {Q^*(S_t, A)}
\label{optimal_pi}
\end{equation}
where $Q*$ can be obtained through the optimal Bellman function that:
\begin{equation}
\begin{aligned}
    Q^*(S_t,A_t) = &\mathbb{E}_{S_{t+1} \sim p_a(\cdot|S_t, A_t)} [R_t + \\ &\gamma \operatorname*{max}\limits_{A} Q^*(S_{t+1}, A)|S_t, A_t]
\end{aligned}
\label{value_function}
\end{equation}

As it is stated while introducing the state transfer probability, $Q^*(S_t, A_t)$ can be approximated through the Mont-Carlo method by updating values in a table $\Tilde{Q}$ that:

\begin{equation}
\begin{aligned}
    \Tilde{Q}\left(S_t, A_t\right) = & \Tilde{Q}\left(S_t, A_t\right) \\
    & +\alpha\left(R_t+\gamma \operatorname*{max}\limits_{A} \Tilde{Q}\left(S_{t+1}, A\right)-\Tilde{Q}\left(S_t, A_t\right)\right)
\end{aligned}
\label{Q_learning}
\end{equation}

where $\gamma$ is a discount factor that indicates the loss of data value over time, which complies with the depreciation that occurs over time in reality; $\alpha \in [0,1]$ is the learning rate and $R_t$ is the system rewards that:
\begin{equation}
    R_t=\sum_{i \in N} r_t^i
\label{Reward}
\end{equation}

\begin{figure}[htbp]
    \centering
    \includegraphics[width=1\linewidth]{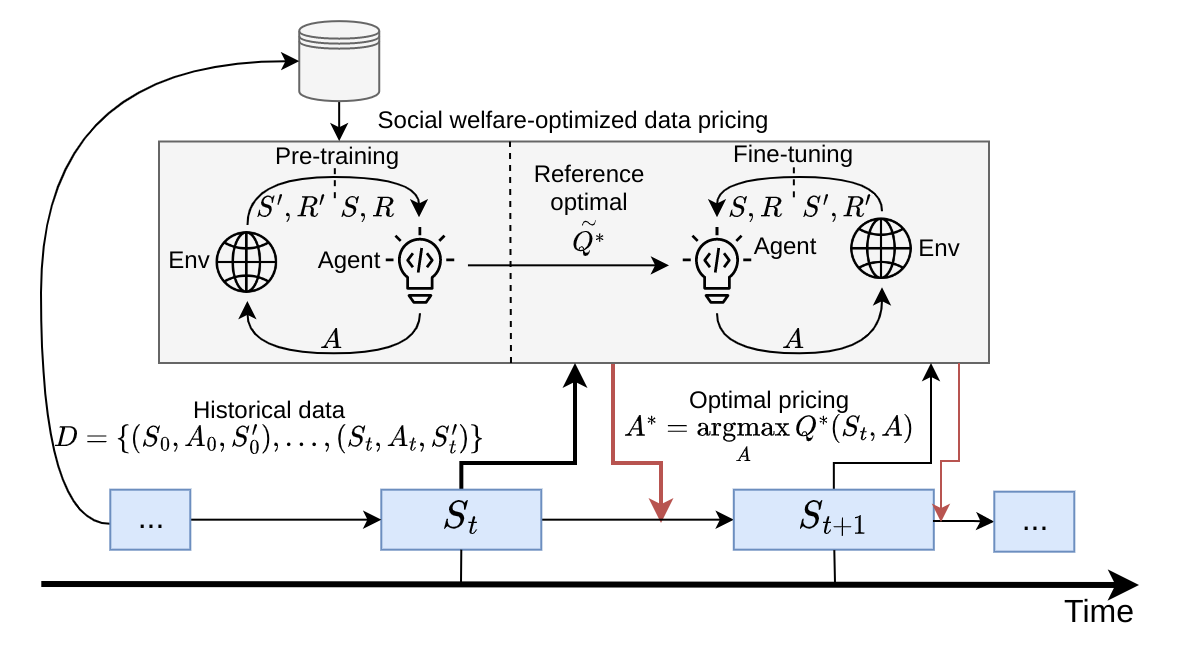}
    \caption{Pre-training and fine-tuning reinforcement learning based data pricing framework: reference optimal $\Tilde{Q'}$ is pre-trained with historic data and passed to fine-tuning for an optimal action collection $A^*$}
    \label{framework}
\end{figure}

The optimal system action-value function can be approximated by the pricing agent in the training process by updating the table $\Tilde{Q}$ through multiple iterations, as shown in Fig. \ref{framework}.


In practical implementations, the efficiency of the pricing algorithm is essential since data trading is real-time. To optimize the pricing process, we could make use of historical data and employ pre-training and fine-tuning strategies, which are proven to reduce the complexity of RL \cite{nairAWACAcceleratingOnline2021,xiePolicyFinetuningBridging2021}. 
Using historical trading data $\mathcal{D}=\{(S_0, A_0, S_0'), ...,(S_t, A_t, S_t')\}$ up to time $t$, we could obtain a reference optimal   $\Tilde{Q'}$. The $\Tilde{Q'}$ is subsequently passed on for further fine-tuning to get optimal $\Tilde{Q}$ and optimal collection of actions $A^*$. The framework of pre-training and fine-tuning strategies is demonstrated in Fig. \ref{framework} while corresponding pseudocodes in Algorithm \ref{algorithm_pricing}. 

\begin{algorithm}[htbp]
\caption{Pre-training and fine-tuning method}\label{algorithm_pricing}
\begin{algorithmic}[1]
\REQUIRE Historic data $\mathcal{D}$, current state $S_t$, learning rate $\alpha$, discount factor $\gamma$, max episodes $\xi$
\ENSURE Optimal action collection $A^*$
\STATE Initialization for pre-training
\FOR{episode $< \xi$}
\STATE Randomly select a sample $(S_t, A_t, S_{t+1})$ from $\mathcal{D}$
\STATE
$\begin{aligned}
\text{Update }  \Tilde{Q'}(S_t, A_t) & = \Tilde{Q'}(S_t, A_t) +\alpha(R_t + \\
& \gamma \operatorname*{max}\limits_{A} \Tilde{Q'}\left(S_{t+1}, A\right)-\Tilde{Q'}(S_t, A_t))
\end{aligned}$
\ENDFOR
\STATE Initialize $S=S_t$, $\Tilde{Q}=\Tilde{Q'}$ for fine-tuning
\WHILE{ $\forall s \in S, v \neq vt$}
\STATE Take $A_t= \operatorname{argmax}\limits_{A} \Tilde{Q}(S_t,A)$, observe $S_{t+1}$ and $R_t$
\STATE 
$\begin{aligned}
\text{Update }  \Tilde{Q}(S_t, A_t) = & \Tilde{Q}(S_t, A_t) +\alpha(R_t + \\
& \gamma \operatorname*{max}\limits_{A} \Tilde{Q}\left(S_{t+1}, A\right)-\Tilde{Q}(S_t, A_t))
\end{aligned}$
\STATE Update $S_t = S_{t+1}$
\ENDWHILE
\RETURN Extract $A^*= \operatorname{argmax}\limits_{A} \Tilde{Q}(S_t,A)$
\end{algorithmic}
\end{algorithm}


The employment of the RL-based pricing mechanism with a matchmaking method could facilitate traders to reach target states through multi-round trading, thereby fully unlocking potential data value and driving the system toward social welfare optimization. 

\subsection{Evaluation Metrics}

The Performance of data pricing mechanisms can be evaluated from feasibility, efficiency, and fairness perspectives.

\paragraph{Feasibility} In the market, rational traders will only accept deals in which the revenue exceeds the cost, which also works in data trading scenarios. A feasible mechanism could optimize the pricing and increase the probability of traders accepting proposals. The feasibility metric is designed based on this:
\begin{equation}
    \Phi_f = M_{a}/M_p
\label{feasibility}
\end{equation}
where $M_{a}$ demonstrates the amount of accomplished actions in testing in one time step; $M$ is the amount of proposals in testing. The computational complexity can be another aspect of feasibility, which can be discussed in the future. 

\paragraph{Efficiency} Generally, an effective pricing mechanism leads to greater revenues per investment for traders and societies, where revenues can be represented by social welfare increments and investment by disclosed information. The efficiency metric is designed based on this:
\begin{equation}
    \Phi_e = W/V
\label{efficiency}
\end{equation}
where $W$ demonstrates the social welfare increment, that $W=\sum_{i \in N} {w^i}$\cite{jiangHighspeedRailPricing2021}; $V$ demonstrate the sum of disclosed information volume, that $V=\sum_{i \in N} dv^i$.

\paragraph{Fairness} Providing equal pay for the same work can motivate traders to cooperate thus bringing higher social welfare increment to the system. The fairness is analyzed through an awareness perspective\cite{dworkFairnessAwareness2011}:

\begin{equation}
    \Phi_r=1-\sqrt{\frac{1}{N} \sum_{i \in N}\left(\phi^{i}-\bar{\phi}\right)^{2}}
\label{fairness}
\end{equation}
where $\bar{\phi}=\frac{1}{N} \sum_{i \in N} \phi^{i}$ and $\phi^i=dc^i/dv^i$ for the trader $i$ in $\mathcal{T}$. All data in this paper is homogeneous, while in future research, we could calculate fairness for the homogeneous data and add all fairness up for system fairness evaluation. 


\section{Numerical simulation}

In this section, we conduct experiments to validate our proposed SWDPM. Due to the lack of real digital asset trading data, numerical simulations are used to evaluate the feasibility and performance.

\subsection{Experiment settings}

The simulation is built on a typical two-sided data trading scenario, as demonstrated in the environment in Fig. \ref{framework}, consisting of four data buyers and suppliers with their preference setting as shown in Table. \ref{parameter_table}. Note that for all trading acceptance intentions, once generated in the initialization, it does not change over time or trading. To evaluate the performance of SWDPM, we conducted four rounds of simulations with different initial states. Traders' initial states $s_0=[vt_0, v_0, c_0]$ are randomly select from the set $\{[10,0,10], [0,10,0], [12,0,12], [0,12,0], [10,0,9]\}$. To reduce the computational complexity, the minimum units $u_v$ and $u_c$ are set to 1, making state space and action space discrete and finite. The hyperparameters for RL training are set as shown in Table. \ref{parameter_table}.

\begin{table}[htbp]
\centering
\caption{Parameter settings in numerical simulations}
\begin{tabular}{cc}
\toprule
Parameter & Value\\
\midrule
$\eta$ & $1$ \\
$\delta$ & $0.2$ \\
$\gamma$ & $0.995$  \\ 
$\alpha$ & $0.1$  \\  
$\theta$ & $-0.5$ \\  
$\lambda$ & $-100$ \\  
$\xi$ & $10^6$ \\  
\bottomrule
\end{tabular}
\label{parameter_table}
\end{table}

To compare with real-world pricing strategies, we employ uniform pricing and subscription-based pricing, which are commonly used in real-world data trading markets\cite{wuOptimalFixedUpToPricing2019, linFullyDecentralizedInfrastructure2020, peiSurveyDataPricing2020}. Uniform pricing is usually used in one-sided markets, in which the price is decided by buyers or sellers based on current market states, while with little consideration of negotiation or sustainability. Subscription-based pricing is usually used in digital product markets, in which the price is fixed and decided by providers concerning long-term revenue, while with little consideration of the dynamics of the data, such as the pricing in Spotify. In these strategies, whether the trading will succeed only depends on another trader's willingness. 

\subsection{Evaluation results} 

\begin{figure}[htbp]
\centering
\includegraphics[width=1\linewidth]{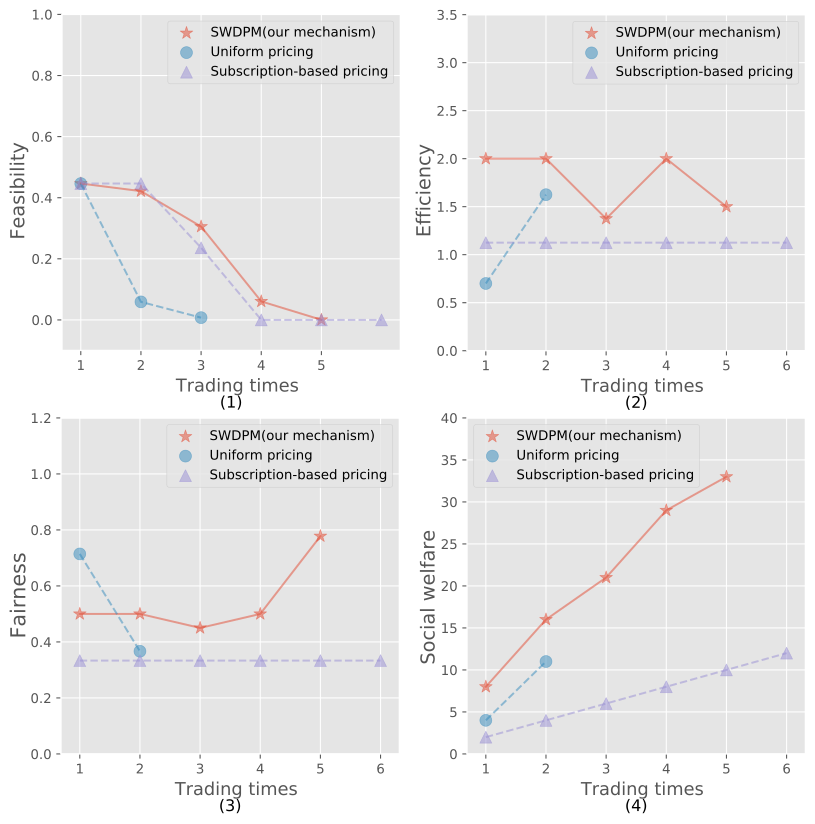}
\caption{Results of evaluation metrics for SWDPM and comparison methods with initial states set as \{$[10, 0, 9]$, [0,10,0]\}}.
\label{result_9_10_1_1_4}
\end{figure}

Evaluation metrics including overall feasibility, efficiency, fairness, and social welfare increment are calculated based on states and actions, as shown in Fig. \ref{result_9_10_1_1_4} and Table. \ref{Evaluation_table}. 

In Fig. \ref{result_9_10_1_1_4}, the horizontal coordinate is the number of trading times, and the vertical coordinate is the evaluation metrics.  As shown in Fig. \ref{result_9_10_1_1_4} (1), SWDPM could greatly increase feasibility because the pricing agent could learn traders' preferences and provide more reasonable pricing through training. The trend of the feasibility curve is consistent with the theoretical trend that as states approach the target state, workable action space becomes smaller and feasibility would naturally decrease. Fig. \ref{result_9_10_1_1_4} (2) shows that efficiency is also promoted in SWDPM, illustrating the overall better revenue for all traders by more reasonable pricing. In Fig. \ref{result_9_10_1_1_4} (3), fairness is not improved in every trading, which is possible because uniform pricing could provide a more fair price considering the randomness. But Fig. \ref{result_9_10_1_1_4} (4) shows that SWDPM could increase the social welfare increment more than 3 times than uniform pricing or subscription-based pricing, which is consistent with our theory that the potential value of data can be fully excavated through multi-round trading. 

Table. \ref{Evaluation_table} shows the average value of evaluation results in one simulation round. More trade times are achieved by SWDPM compared with uniform pricing, which means the characteristics of trading and preferences of traders are captured through training, thus the trading could be continually incentivized. Other evaluation results in table.\ref{Evaluation_table} also align with Fig. \ref{result_9_10_1_1_4}, indicating the robustness of SWDPM that it can stably motivate data trading and achieve social welfare optimization in all simulation rounds.

Overall, simulation results indicate that it is essential to consider the data's dynamic characteristics under the condition of multi-round trading. Using matchmaking-based MDP and setting social welfare optimization objectives in the RL, SWDPM could capture the dynamics during the trading process and achieve social welfare optimization by sustainably incentivizing the trading process.

\begin{table}[htbp]
\centering
\caption{Results of evaluation metrics for simulations}
\begin{tabular}{ccccccc}
\toprule
\multicolumn{1}{l}{} & \multicolumn{1}{l}{Methods} & Times & $\Phi_f$ &  $\Phi_e$ &  $\Phi_r$
 &$\sum W$ \\
\midrule
\multirow{2}{*}{Round1} & \textbf{SWDPM} & \textbf{5} & 0.261 & 1.8 & 0.533 & \textbf{34} \\
                        & Uniform & 2 & 0.313 & 1.875 & 0.625 & 13 \\
\multirow{2}{*}{Round2} & \textbf{SWDPM}  & \textbf{5}  & 0.235 & 1.875 & 0.533 & \textbf{35} \\
                        & Uniform  & 3 & 0.188 & 1.34 & 0.687 & 17 \\
\multirow{2}{*}{Round3} & \textbf{SWDPM} & \textbf{6} & 0.094 & 1.896 & 0.528 & \textbf{43} \\
                        & Uniform & 2 & 0.061 & 1.76 & 0.706 & 13 \\
\multirow{2}{*}{Round4} & \textbf{SWDPM} & \textbf{6} & 0.094 & 1.917 & 0.528 & \textbf{43} \\
                        & Uniform  & 2 & 0.092 & 1.333 & 0.539 & 12 \\
\bottomrule
\end{tabular}
\label{Evaluation_table}
\end{table}

\section{Conclusion and future works}
In this paper, we propose a novel mechanism, SWDPM, that supports multi-round trading patterns. The SWDPM models trading as a matchmaking-based MDP and utilizes pre-training and fine-tuning RL to optimize pricing strategies. Compared with uniform and subscription-based pricing, numerical simulations demonstrate that SWDPM can increase social welfare by encouraging better matching of demand and price negotiation among traders, improving feasibility, efficiency, and fairness.

Our proposed mechanism is not only feasible but also social welfare-optimized for multi-round trading patterns, which could provide theoretical support for the development of healthy and sustainable data trading markets by offering fair and reasonable pricing. Considering further improvement and extension, some work awaits future research from the community. For example, in theoretical modeling, temporal correlations between multiple data pricing actions may exist, requiring the use of semi-MDP to bundle related states into one state, enhancing the model's accuracy. In addition, the training process in SWDPM could be decentralized to reduce the computing burden and facilitate implementation in broader markets with a critical mass of traders\cite{liFutureManagementDAO2022}. Further, for practical applications, smart contracts and platform service fees in Blockchain research offer insights to enhance the practicality and efficiency of SWDPM\cite{liAnalyzingBitcoinTransaction2022, liTransactionQueuingGame2018}, as matchmaking can be computationally expensive and difficult to accomplish automatically and freely in reality.

\section*{Acknowledgment}
This work is supported by the Shanghai Artificial Intelligence Laboratory, the Science and Technology Development Fund, Macau, SAR (0050/2020/A1), and the National Natural Science Foundation of China (62103411).



\begin{thebibliography}{10}
\providecommand{\url}[1]{#1}
\csname url@samestyle\endcsname
\providecommand{\newblock}{\relax}
\providecommand{\bibinfo}[2]{#2}
\providecommand{\BIBentrySTDinterwordspacing}{\spaceskip=0pt\relax}
\providecommand{\BIBentryALTinterwordstretchfactor}{4}
\providecommand{\BIBentryALTinterwordspacing}{\spaceskip=\fontdimen2\font plus
\BIBentryALTinterwordstretchfactor\fontdimen3\font minus
  \fontdimen4\font\relax}
\providecommand{\BIBforeignlanguage}[2]{{%
\expandafter\ifx\csname l@#1\endcsname\relax
\typeout{** WARNING: IEEEtran.bst: No hyphenation pattern has been}%
\typeout{** loaded for the language `#1'. Using the pattern for}%
\typeout{** the default language instead.}%
\else
\language=\csname l@#1\endcsname
\fi
#2}}
\providecommand{\BIBdecl}{\relax}
\BIBdecl

\bibitem{zhengServiceGeneratedBigData2013}
Z.~Zheng, J.~Zhu, and M.~R. Lyu, ``Service-generated big data and big
  data-as-a-service: An overview,'' in \emph{2013 IEEE International Congress
  on Big Data}, Jun. 2013, pp. 403--410.

\bibitem{insights_big_2022}
\BIBentryALTinterwordspacing
F.~B. Insights, ``\BIBforeignlanguage{en}{Big {Data} {Analytics} {Market}
  {Size} [2022-2029] {Exhibits} 13.4\% {CAGR} to {Reach} {USD} 655.53 {Billion}
  in 2029},'' Nov. 2022. [Online]. Available:
  \url{https://www.globenewswire.com/en/news-release/2022/11/14/2554626/0/en/Big-Data-Analytics-Market-Size-2022-2029-Exhibits-13-4-CAGR-to-Reach-USD-655-53-Billion-in-2029.html}
\BIBentrySTDinterwordspacing

\bibitem{tripathiDoesInformationCommunications2020}
M.~Tripathi and S.~K. Inani, ``Does information and communications technology
  affect economic growth? empirical evidence from saarc countries,''
  \emph{Information Technology for Development}, vol.~26, no.~4, pp. 773--787,
  Oct. 2020.

\bibitem{labCityDataExchange2017}
C.~S. Lab, ``City data exchange copenhagen,''
  https://cphsolutionslab.dk/en/projekter/data-platforms/city-data-exchange,
  2017.

\bibitem{agarwalMarketplaceDataAlgorithmic2019}
A.~Agarwal, M.~Dahleh, and T.~Sarkar, ``A marketplace for data: An algorithmic
  solution,'' May 2019.

\bibitem{zhangSurveyDataPricing2020}
M.~Zhang and F.~Beltr{\'a}n, ``A survey of data pricing methods,'' Rochester,
  NY, Apr. 2020.

\bibitem{chenSellingDataMachine2022}
J.~Chen, M.~Li, and H.~Xu, ``Selling data to a machine learner: Pricing via
  costly signaling,'' in \emph{Proceedings of the 39th International Conference
  on Machine Learning}.\hskip 1em plus 0.5em minus 0.4em\relax PMLR, Jun. 2022,
  pp. 3336--3359.

\bibitem{yuDataPricingStrategy2017}
H.~Yu and M.~Zhang, ``Data pricing strategy based on data quality,''
  \emph{Computers \& Industrial Engineering}, vol. 112, pp. 1--10, Oct. 2017.

\bibitem{tianDataBoundaryData2022}
Y.~Tian, Y.~Ding, S.~Fu, and D.~Liu, ``Data boundary and data pricing based on
  the shapley value,'' \emph{IEEE Access}, vol.~10, pp. 14\,288--14\,300, 2022.

\bibitem{koutrisQueryBasedDataPricing2015}
P.~Koutris, P.~Upadhyaya, M.~Balazinska, B.~Howe, and D.~Suciu, ``Query-based
  data pricing,'' \emph{Journal of the ACM}, vol.~62, no.~5, pp. 43:1--43:44,
  Nov. 2015.

\bibitem{peiSurveyDataPricing2020}
J.~Pei, ``A survey on data pricing: From economics to data science,''
  \emph{IEEE Transactions on Knowledge and Data Engineering}, pp. 1--1, 2020.

\bibitem{niuMakingBigMoney2019}
C.~Niu, Z.~Zheng, S.~Tang, X.~Gao, and F.~Wu, ``Making big money from small
  sensors: Trading time-series data under pufferfish privacy,'' in \emph{IEEE
  INFOCOM 2019 - IEEE Conference on Computer Communications}, Apr. 2019, pp.
  568--576.

\bibitem{zhouBriefSurveyAnonymization2008}
B.~Zhou, J.~Pei, and W.~Luk, ``A brief survey on anonymization techniques for
  privacy preserving publishing of social network data,'' \emph{ACM SIGKDD
  Explorations Newsletter}, vol.~10, no.~2, pp. 12--22, Dec. 2008.

\bibitem{liTheoryPricingPrivate2015}
C.~Li, D.~Y. Li, G.~Miklau, and D.~Suciu, ``A theory of pricing private data,''
  \emph{ACM Transactions on Database Systems}, vol.~39, no.~4, pp. 34:1--34:28,
  Dec. 2015.

\bibitem{mothukuriSurveySecurityPrivacy2021}
V.~Mothukuri, R.~M. Parizi, S.~Pouriyeh, Y.~Huang, A.~Dehghantanha, and
  G.~Srivastava, ``A survey on security and privacy of federated learning,''
  \emph{Future Generation Computer Systems}, vol. 115, pp. 619--640, Feb. 2021.

\bibitem{liNovelFrameworkData2022}
C.~Li, Y.~Yuan, and F.-Y. Wang, ``A novel framework for data trading markets
  based on blockchain-enabled federated learning,'' in \emph{2022 IEEE 25th
  International Conference on Intelligent Transportation Systems (ITSC)}, Oct.
  2022, pp. 3392--3397.

\bibitem{changIncorporatingMarkovDecision2017}
Y.-H. Chang and M.-S. Lee, ``Incorporating markov decision process on genetic
  algorithms to formulate trading strategies for stock markets,'' \emph{Applied
  Soft Computing}, vol.~52, pp. 1143--1153, Mar. 2017.

\bibitem{jiangHighspeedRailPricing2021}
C.~Jiang and C.~Wang, ``High-speed rail pricing: Implications for social
  welfare,'' \emph{Transportation Research Part E: Logistics and Transportation
  Review}, vol. 155, p. 102484, Nov. 2021.

\bibitem{wuOptimalFixedUpToPricing2019}
S.~Wu and P.~Pavlou, ``On the optimal fixed-up-to pricing for information
  services,'' \emph{Journal of the Association for Information Systems},
  vol.~20, no.~10, Oct. 2019.

\bibitem{linFullyDecentralizedInfrastructure2020}
C.-H.~V. Lin, C.-C.~J. Huang, Y.-H. Yuan, and Z.-s.~S. Yuan, ``A fully
  decentralized infrastructure for subscription-based iot data trading,'' in
  \emph{2020 IEEE International Conference on Blockchain (Blockchain)}, Nov.
  2020, pp. 162--169.

\bibitem{liNovelGSPAuction2022}
J.~Li, X.~Ni, Y.~Yuan, and F.-Y. Wang, ``A novel gsp auction mechanism for
  dynamic confirmation games on bitcoin transactions,'' \emph{IEEE Transactions
  on Services Computing}, vol.~15, no.~3, pp. 1436--1447, May 2022.

\bibitem{liNovelGSPAuction2019}
J.~Li, Y.~Yuan, and F.-Y. Wang, ``A novel gsp auction mechanism for ranking
  bitcoin transactions in blockchain mining,'' \emph{Decision Support Systems},
  vol. 124, p. 113094, Sep. 2019.

\bibitem{mehtaHowSellDataset2019}
S.~Mehta, M.~Dawande, G.~Janakiraman, and V.~Mookerjee, ``How to sell a
  dataset? pricing policies for data monetization: 20th acm conference on
  economics and computation, ec 2019,'' \emph{ACM EC 2019 - Proceedings of the
  2019 ACM Conference on Economics and Computation}, p. 679, Jun. 2019.

\bibitem{fernandezDataMarketPlatforms2020}
R.~C. Fernandez, P.~Subramaniam, and M.~J. Franklin, ``Data market platforms:
  Trading data assets to solve data problems,'' \emph{Proceedings of the VLDB
  Endowment}, vol.~13, no.~12, pp. 1933--1947, Aug. 2020.

\bibitem{maestreReinforcementLearningFair2019}
R.~Maestre, J.~Duque, A.~Rubio, and J.~Arevalo, ``Reinforcement learning for
  fair dynamic pricing,'' in \emph{Intelligent Systems and Applications}, ser.
  Advances in Intelligent Systems and Computing, K.~Arai, S.~Kapoor, and
  R.~Bhatia, Eds.\hskip 1em plus 0.5em minus 0.4em\relax Cham: Springer
  International Publishing, 2019, pp. 120--135.

\bibitem{kongOnlinePricingDemand2020}
X.~Kong, D.~Kong, J.~Yao, L.~Bai, and J.~Xiao, ``Online pricing of demand
  response based on long short-term memory and reinforcement learning,''
  \emph{Applied Energy}, vol. 271, p. 114945, Aug. 2020.

\bibitem{xuDeepReinforcementLearning2019}
H.~Xu, H.~Sun, D.~Nikovski, S.~Kitamura, K.~Mori, and H.~Hashimoto, ``Deep
  reinforcement learning for joint bidding and pricing of load serving
  entity,'' \emph{IEEE Transactions on Smart Grid}, vol.~10, no.~6, pp.
  6366--6375, Nov. 2019.

\bibitem{krasheninnikovaReinforcementLearningPricing2019}
E.~Krasheninnikova, J.~Garc{\'i}a, R.~Maestre, and F.~Fern{\'a}ndez,
  ``Reinforcement learning for pricing strategy optimization in the insurance
  industry,'' \emph{Engineering Applications of Artificial Intelligence},
  vol.~80, pp. 8--19, Apr. 2019.

\bibitem{shenReinforcementMechanismDesign2020}
W.~Shen, B.~Peng, H.~Liu, M.~Zhang, R.~Qian, Y.~Hong, Z.~Guo, Z.~Ding, P.~Lu,
  and P.~Tang, ``Reinforcement mechanism design: With applications to dynamic
  pricing in sponsored search auctions,'' \emph{Proceedings of the AAAI
  Conference on Artificial Intelligence}, vol.~34, no.~02, pp. 2236--2243, Apr.
  2020.

\bibitem{harrisMakertakerPricingEffects2013}
L.~Harris, ``Maker-taker pricing effects on market quotations,'' \emph{USC
  Marshall School of Business Working Paper. Avalable at http://bschool. huji.
  ac. il/. upload/hujibusiness/Maker-taker. pdf}, 2013.

\bibitem{nairAWACAcceleratingOnline2021}
A.~Nair, A.~Gupta, M.~Dalal, and S.~Levine, ``Awac: Accelerating online
  reinforcement learning with offline datasets,'' Apr. 2021.

\bibitem{xiePolicyFinetuningBridging2021}
T.~Xie, N.~Jiang, H.~Wang, C.~Xiong, and Y.~Bai, ``Policy finetuning: Bridging
  sample-efficient offline and online reinforcement learning,'' \emph{Advances
  in neural information processing systems}, vol.~34, pp. 27\,395--27\,407,
  2021.

\bibitem{dworkFairnessAwareness2011}
C.~Dwork, M.~Hardt, T.~Pitassi, O.~Reingold, and R.~Zemel, ``Fairness through
  awareness,'' Nov. 2011.

\bibitem{liFutureManagementDAO2022}
J.~Li, R.~Qin, and F.-Y. Wang, ``The future of management: Dao to smart
  organizations and intelligent operations,'' \emph{IEEE Transactions on
  Systems, Man, and Cybernetics: Systems}, pp. 1--11, 2022.

\bibitem{liAnalyzingBitcoinTransaction2022}
J.~Li, Y.~Yuan, and F.-Y. Wang, ``Analyzing bitcoin transaction fees using a
  queueing game model,'' \emph{Electronic Commerce Research}, vol.~22, no.~1,
  pp. 135--155, Mar. 2022.

\bibitem{liTransactionQueuingGame2018}
J.~Li, Y.~Yuan, S.~Wang, and F.-Y. Wang, ``Transaction queuing game in bitcoin
  blockchain,'' in \emph{2018 IEEE Intelligent Vehicles Symposium (IV)}, Jun.
  2018, pp. 114--119.

\end{thebibliography}

\end{document}